\documentclass{appolb}%
\usepackage{amssymb}
\usepackage{epsfig}
\usepackage{amsmath}
\usepackage{graphicx}
\usepackage{epsfig}
\usepackage{amsmath}
\usepackage{amssymb}
\usepackage{amsfonts}%

\begin{document}

\title{Importance of Mixing for Exotic Baryons}
\author{Michal Praszalowicz \address{
M. Smoluchowski Institute of Physics,
Jagellonian University,
Reymonta 4, 30-059 Krakow, Poland}}
\maketitle

\begin{abstract}
Exotic antidecuplet baryons are predicted to be not only surprisingly light
but also very narrow. First we explain how small decay width arises in the
quark soliton model. Next, we study possible mixing of exotic antidecuplet
with Roper octet and discuss its phenomenological consequences.

\end{abstract}


\section{Introduction}

\label{intro}

Despite recent scepticism concerning early announcements of the discovery of
exotic strange baryon $\Theta^{+}(1540)$ two collaborations DIANA and LEPS
confirmed their original results
\cite{Nakano_confirm,Hotta:2005rh,Barmin_confirm}. We refer the reader to
recent experimental reviews
\cite{Nakano:2005ut,Burkert:2005ft,Danilov:2008zza}. In this article following
\cite{Goeke:2009ae} we assume that $\Theta^{+}$ exists with a mass equal 1540
MeV and total width $\Gamma<1$ MeV. An immediate consequences is the existence
of the whole exotic SU(3) multiplet: $\overline{10}$. Apart from truly
exotic states antidecuplet contains cryptoexotic nucleon- ($N_{\overline{10}}%
$) and $\Sigma$-like states ($\Sigma_{\overline{10}}$). The interpretation of
these states is not well understood: one may try to associate them with some
known resonances, or one may postulate the existence of new resonances with
nucleon or $\Sigma$ quantum numbers. This is the approach which we adopt here
\cite{Goeke:2009ae}. Following Refs.~\cite{Arndt:2003ga}%
\nocite{Kuznetsov04,gra1,Kuznetsov:2008hj}--\cite{Kuznetsov:2010as} we assume that there
exists new, narrow nucleon resonance $N(1685)$ which we will interpret as a
member of $\overline{10}$. If so, $N_{\overline{10}}$ decays have to satisfy
the following constraints \cite{Goeke:2009ae} originally discussed in \cite{Arndt:2003ga}:%
\begin{equation}
\Gamma_{N_{\overline{10}}\rightarrow\pi N}<0.5\;\text{MeV, }\;\text{Br}%
_{N_{\overline{10}}\rightarrow\eta N}>0.2,\;5\;\text{MeV}<\Gamma
_{N_{\overline{10}}}^{\text{tot}}<25\;\text{MeV.} \label{conditions}%
\end{equation}
Unfortunately the small partial width of $N(1685)$ to $\pi N$ contradicts
SU(3) symmetry relations between the decay constants of $N_{\overline{10}}$.

In this short note we would like to emphasize the importance of mixing both
for decays and mass spectra of the putative exotic antidecuplet baryons. The
exotic states have been already anticipated by the founders of the quark
model although they did not elaborate on them. Later the arguments have
been raised that they should be heavy and wide. In contrast, chiral
soliton models predicted that the pentaquark masses were
generically small (\emph{i.e.} in the range of 1.5--1.6 GeV)
\cite{BieDot,prasz,DPP}. It was much more difficult to accommodate the small
decay width of $\Theta^{+}$ \cite{Weigel}. In Sect.~\ref{decaywidths} we
explain how small decay width arises naturally in the Chiral Soliton Quark
Model ($\chi$QSM). Next, in Sect.~\ref{GMO} we argue that the decay coupling
$g_{\Theta NK}$ is further reduced due to Gell-Mann--Okubo (GMO) mixing
caused by
the nonzero $m_{s}$. In Sect.~\ref{Rmixing} we show how additional mixing of
$\overline{10}$ with Roper octet can change decay patterns of $N_{\overline
{10}}$. We estimate allowed range of mixing angles and present predictions for
remaining members of $\overline{10}$: $\Sigma_{\overline{10}}$ and
$\Xi_{\overline{10}}$. Conclusions are presented in Sect.~\ref{summ}.

\section{Decay widths in Chiral Soliton Quark Model}

\label{decaywidths}

In in Ref.~ \cite{DPP}  the
following nonrelativistic formula for the decay width has been used:
\begin{equation}
\Gamma_{B_{1}\rightarrow B_{2}\varphi}=\frac{g_{B_{1}B_{2}\varphi}^{2}}%
{2\pi(M_{1}+M_{2})^{2}} \, p_{\varphi}^{3}. \label{Gammadef}%
\end{equation}
It follows from the Goldberger-Treimann relation between axial and strong decay constants. Here $M_{1}$ is the mass of the decaying baryon, $M_{2}$ the mass of the decay product and $p_{\varphi}$ is the outgoing meson momentum.
Generically $\Gamma_{\Theta^+\rightarrow NK}$ given by (\ref{Gammadef})
would be still in the range of a few hundreds of MeV \cite{Weigel} if not for the terms non-leading in $1/N_{c}$ expansion. Indeed, even without mixing
the decay constant $g_{B_{1}B_{2}\varphi}$, which stands for the matrix
element of the tensor decay operator $O_{\varphi}^{(8)}$ between
the physical baryon wave functions $\left\vert B^{\text{phys}}\right\rangle $:%
\begin{equation}
g_{B_{1}B_{2}\varphi}=\left\langle B_{2}^{\text{phys}}\right\vert O_{\varphi
}^{(8)}\left\vert B_{1}^{\text{phys}}\right\rangle \label{gdef}%
\end{equation}
is the
sum of three different contributions that are formally of different order in
$1/N_{c}$. However, they are multiplied by the SU(3) Clebsch-Gordan
coefficients that also depend on $N_{c}$ \cite{Praszalowicz:2003tc}. For
example:
\begin{equation}
g_{\Theta NK}^{2}=\frac{9(N_{c}+1)}{(N_{c}+3)(N_{c}+7)}G_{\overline{10}}%
^{2}\quad\text{with\quad}G_{\overline{10}}=G_{0}-\frac{N_{c}+1}{4}G_{1}%
-\frac{1}{2}G_{2} \label{gTNK}%
\end{equation}
where $G_{0}\sim N_{c}^{3/2},\,G_{1,2}\sim N_{c}^{1/2}$. Similarly%
\begin{equation}
g_{\Delta N\pi}^{2}=\frac{3(N_{c}-1)(N_{c}+5)}{2(N_{c}+1)(N_{c}+7)}G_{10}%
^{2}\quad\text{with}\quad G_{10}=G_{0}+\frac{1}{2}G_{2}. \label{gDNpi}%
\end{equation}
Chiral soliton models provide us with specific predictions for constants
$G_{0,1,2}$ \cite{Ellis:2004uz}. Had we neglected $G_{1}$ and $G_{2}$ in
Eqs.(\ref{gTNK},\ref{gDNpi}) (which would be inconsistent for $g_{\Theta NK}$
because of $N_{c}$ enhancement of $G_{1}$) we would have obtained for
$N_{c}=3$:%
\[
g_{\Delta N\pi}=g_{\Theta NK}\sim17.6
\]
estimating $G_{0}$ from the experimental value of $\Delta$ decay width, and
consequently%
\begin{equation}
\Gamma_{\Theta NK}\sim150\;\text{MeV.}%
\end{equation}
We see that small decay width of $\Theta^{+}$ results from the cancelation in
(\ref{gTNK}). Indeed, the authors of Ref. \cite{DPP} have shown that in the
nonrelativistic limit of $\chi$QSM one obtains that $G_{0}=-(N_{c}+2)\,G,\quad
G_{1}=-4G,\quad G_{2}=-2G,$ with $G\sim N_{c}^{1/2} $ and consequently
$\Gamma_{\Theta NK}=0$! In the same limit $\chi$QSM predicts that $g_{A}=3/5$
and $\mu_{p}/\mu_{n}=-3/2$. It follows that antidecuplet decay constants are small.

\section{Gell-Mann Okubo Mixing}

\label{GMO}

Treating $m_{s}$ corrections as perturbation introduces mixing
\cite{Goeke:2009ae}%
\begin{align}
\left\vert B_{8}^{\text{phys}}\right\rangle  &  =\left\vert 8_{1/2}%
,B\right\rangle +c_{\overline{10}}^{B}\left\vert \overline{10}_{1/2}%
,B\right\rangle +c_{27}^{B}\left\vert 27_{1/2},B\right\rangle \,,\nonumber\\
\left\vert B_{\overline{10}}^{\text{phys}}\right\rangle  &  =\left\vert
\overline{10}_{1/2},B\right\rangle +d_{8}^{B}\left\vert 8_{1/2},B\right\rangle
+d_{27}^{B}\left\vert 27_{1/2},B\right\rangle +d_{\overline{35}}^{B}\left\vert
\overline{35}_{1/2},B\right\rangle \, \label{admix}%
\end{align}
where subscripts refer to spin. For some specific states mixing constants
$c_{R}^{B},\,d_{R}^{B}\sim m_{s}$ may be equal zero due to the isospin. For
example $\Theta^{+}$ mixes only with $\overline{35}$, but this component of
the wave function does not contribute to the decays to octet. Therefore only
mixing of the final nucleon with $\overline{10}$ and $27$ modifies
the decay constant:%
\begin{equation}
g_{\Theta NK}^{2}=\frac{3}{5}\left[  G_{\overline{10}}+\frac{5}{4}%
c_{\overline{10}}\,H_{\overline{10}}-\frac{7}{4}c_{27}H_{27}^{\prime}\right]
^{2}. \label{gTNKmix}%
\end{equation}
Since $G_{\overline{10}}$ is small the admixtures proportional to the reduced
matrix elements%
\begin{equation}
\,H_{\overline{10}}\sim\left\langle \overline{10}_{1/2},B^{\prime}\right\vert
\hat{O}_{\varphi}^{(8)}\left\vert \overline{10}_{1/2},B\right\rangle ,\quad
H_{27}^{\prime}\sim\left\langle 27_{1/2},B^{\prime}\right\vert \hat
{O}_{\varphi}^{(8)}\left\vert \overline{10}_{1/2},B\right\rangle
\end{equation}
are important even if mixing parameters $c_{\overline{10}}$ and $c_{27}$ are
not large (for definitions see Ref.~\cite{Praszalowicz:2004dn}). Neither
$H_{\overline{10}}$ nor $H_{27}^{\prime}$ vanish in the nonrelativistic limit.
Therefore in this limit $\Theta^{+}$ decay occurs entirely due to the mixing.
For realistic model parameters (when $G_{\overline{10}}>0$) there is a
cancelation (note that $H_{\overline{10}}<0$ and $H_{27}^{\prime}>0$) between
different terms in (\ref{gTNKmix}) and $g_{\Theta NK}$ is further suppressed.
Mixing affects decay patterns of pentaquarks violating SU(3) relations between
the decay constants $g_{B_{1}B_{2}\varphi}$ \cite{Praszalowicz:2004dn}.

On somewhat more phenomenological ground let us consider only
$8\leftrightarrow\overline{10}$ mixing \cite{Goeke:2009ae} defining%
\begin{equation}
g_{\theta NK}=\cos\alpha\,g_{\overline{10}}+\sin\alpha\,h_{\overline{10}}%
\end{equation}
which can be directly extracted from (\ref{Gammadef}) if $\Gamma_{\Theta NK} $
is known. Throughout this paper we assume that $\Gamma_{\Theta NK}\simeq1$
MeV, hence $g_{\theta NK}\simeq1.4$. It is then possible to express decay
constants of other members of antidecuplet in terms of measurable physical
parameters such as $g_{\pi NN}\simeq13.2,\,\varepsilon=F/D\simeq0.56,$ mixing
angle $\alpha$ and one a priori unknown parameter $h_{\overline{10}}$ which
can be estimated from $\chi$QSM calculations. Here we take $h_{\overline{10}%
}=-7$ \cite{Goeke:2009ae}. Then we have for example:%
\begin{align}
g_{N_{\overline{10}}N\pi}  &  =\frac{1}{2}\cos\alpha\,g_{\theta NK}%
-\,\tan\alpha\sqrt{3}g_{\pi NN},\nonumber\\
g_{N_{\overline{10}}N\eta}  &  =\frac{1}{2}\cos\alpha\,g_{\theta NK}-\frac
{1}{2}\sin2\alpha\,h_{\overline{10}}+\tan\alpha\frac{3\varepsilon
-1}{1+\varepsilon}\frac{g_{\pi NN}}{\sqrt{3}}.
\end{align}


If we want to interpret $N(1685)$ as $N_{\overline{10}}$ we need to satisfy
bounds (\ref{conditions}). This is quite difficult within one angle scenario.
It is possible to nullify $g_{N_{\overline{10}}N\pi}$ by a suitable choice of
mixing angle $\alpha\sim0.03$,
but then the mean octet mass (i.e. the nucleon mass before mixing) which is
experimentally $1151$ MeV comes out wrong \cite{Goeke:2009ae}:%
\begin{equation}
M_{8}=M_{N}^{\text{phys}}\cos^{2}\alpha+M_{N_{\overline{10}}}^{2}\sin
^{2}\alpha\simeq M_{N}^{\text{phys}}\;.
\end{equation}
The mixing angle that satisfies (\ref{conditions}) is order of magnitude too
small to account for baryon masses. For realistic mixing angles
$g_{N_{\overline{10}}N\pi}$ is dominated by $g_{\pi NN}$ and $\Gamma
_{N_{\overline{10}}N\pi}>\Gamma_{N_{\overline{10}}N\eta}$ in contradiction
with experimental data for $N(1685)$.

\section{Mixing with Roper}

\label{Rmixing}

\begin{figure}[t]
\begin{centering}
\includegraphics[scale=0.6]{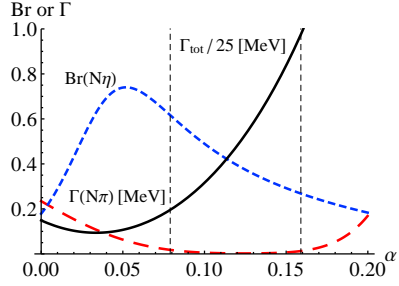}
\par\end{centering}
\caption{Total (solid black line) decay width (divided by 25) and partial
decay width of $N_{\overline{10}}$ to $\pi N$ (long dashed red line) in MeV
together with branching ratio (short dashed blue line) of $N_{\overline{10}%
}\rightarrow\eta N$ as functions of mixing angle $\alpha$ along the line
(\ref{or_lin}). Thin vertical lines correspond to the limits on the mixing
angle $\alpha$. The plot is made for $h_{\overline{10}}=-7$.}%
\label{fig:twoangles}%
\end{figure}

In order to satisfy conditions (\ref{conditions}) with more realistic mixing
angles we have considered in Ref.~\cite{Goeke:2009ae} scenario in which exotic
antidecuplet can mix with Roper resonance octet (mixing angle $\phi$). For
Roper octet GMO mass formulae work with much worse accuracy than for the
ground state octet \cite{dia1,guz2}, so there is a need for additional mixing.
Since $\varepsilon_{\text{Roper}}\sim1/3$ \cite{guz2} Roper decay to $N\eta$
is negligible. However Roper admixture contributes to $N\pi$ and other decay
modes:%
\begin{align}
g_{N_{\overline{10}}N\pi}  &  =\frac{1}{2}\cos\phi\cos\alpha\,g_{\theta
NK}-\,\cos\phi\tan\alpha\sqrt{3}g_{\pi NN}-\tan\phi\,g_{RN\pi},\label{gN10bar}%
\\
g_{N_{\overline{10}}N\eta}  &  =\frac{1}{2}\cos\phi\cos\alpha\,g_{\theta
NK}-\frac{1}{2}\cos\phi\sin2\alpha\,h_{\overline{10}}+\cos\phi\tan\alpha
\frac{3\varepsilon-1}{1+\varepsilon}\frac{g_{\pi NN}}{\sqrt{3}}.\nonumber
\end{align}
Since $g_{RN\pi}\sim12$ is comparable with $g_{\pi NN}$ one may suppress
$g_{N_{\overline{10}}N\pi}$ without changing much $g_{N_{\overline{10}}N\eta}%
$. In Ref.~\cite{Goeke:2009ae} we have found that conditions (\ref{conditions}%
) are satisfied in the vicinity of the line
\begin{equation}
\phi(\alpha)=0.0508-2.207\alpha,\quad0.079<\alpha<0.159. \label{or_lin}%
\end{equation}
We see that mixing angles are reasonable. The decay widths and branching ratio
to $N\eta$ along (\ref{or_lin}) are plotted in Fig.~\ref{fig:twoangles}

\renewcommand{\baselinestretch}{0.5} \renewcommand{\baselinestretch}{1.0}

\begin{figure}[t]
\begin{centering}
\includegraphics[scale=0.65]{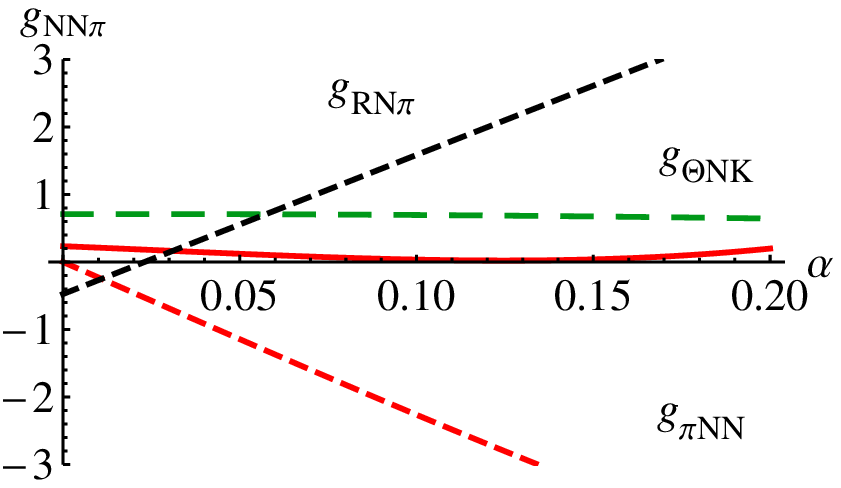}
\hspace{0.3cm}
\includegraphics[scale=0.65]{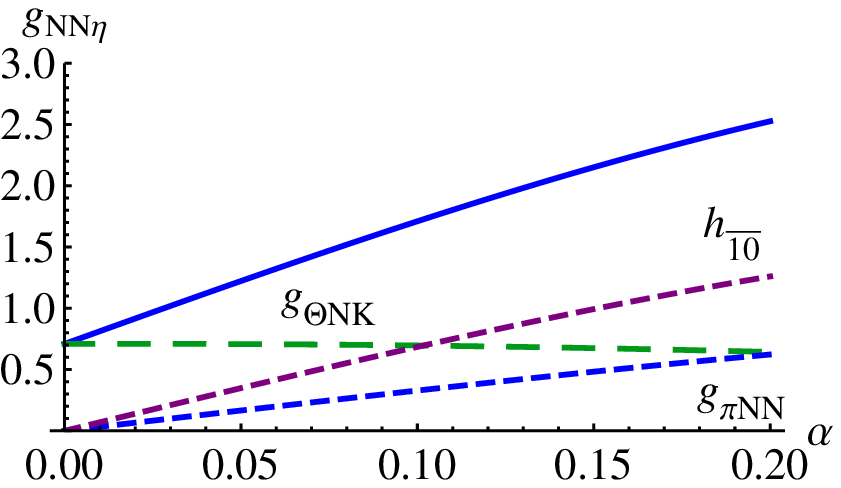}
\par\end{centering}
\caption{Decay constants of $N_{\overline{10}}$ (solid lines) with different
components defined in Eq.~(\ref{gN10bar}) shown by dashed lines. The plot is
made for $h_{\overline{10}}=-7$.}%
\label{fig:gN10bar}%
\end{figure}

In Fig.~\ref{fig:gN10bar} we plot $g_{N_{\overline{10}}N\pi}$ and
$g_{N_{\overline{10}}N\eta} $ together with their different components along
the line (\ref{or_lin}). We see that indeed $g_{N_{\overline{10}}N\pi}$ is
small due to the cancellation between $g_{\pi NN}$ and $g_{RN\pi}$, while
$g_{N_{\overline{10}}N\eta}$ rises moderately when mixing increases.

\renewcommand{\baselinestretch}{0.5} \renewcommand{\baselinestretch}{1.0}

Having established the range of mixing angles we can predict masses of
the remaining antidecuplet members \cite{Goeke:2009ae}:%
\begin{align}
1795\;\text{MeV}  &  <M_{\Sigma_{\overline{10}}}<1830\;\text{MeV,}%
\label{sigmarange}\\
1900\;\text{MeV}  &  <M_{\Xi_{\overline{10}}}<1970\;\text{MeV}. \label{xirange}%
\end{align}
Note that bound (\ref{xirange}) contradicts the result of NA49 $M_{\Xi
_{\overline{10}}}\sim1860$~MeV \cite{Alt:2003vb}. From our analysis it follows
that total decay width of $\Xi_{\overline{10}}$ to $K\Sigma$ and $\pi\Xi$ is
of the order of 10 MeV. Total width of $\Sigma_{\overline{10}}$ does not
exceed 30 MeV but is also constrained from below to be larger that 10 MeV.
Most prominent decay channels are $KN$ and $\pi\Lambda$ with branching ratios
approximately 60\% and 20\% respectively. Due to the mixing SU(3) forbidden
decays to decuplet are possible, but small, at the level of 5 to 9\%.

\section{Summary and Conclusions}

\label{summ}

Mixing induced by $m_{s}$ was first studied within the $\chi$QSM already in
Ref. \cite{DPP} but only in the leading order in $N_{c}$. It was extended to
nonleading terms in Ref.~\cite{Arndt:2003ga} and \cite{Praszalowicz:2004dn}.
Mixing apears also in other approaches to pentaquarks. For example in a
diquark model \cite{JafWil} antidecuplet mixes with an accompanying
cryptoexotic octet. Diagonalization of strangeness induces \emph{ideal}
(large) mixing between these two representations. The resulting physical
states of nucleon quantum numbers have been interpreted as Roper and $N^{\ast
}(1710)$. However due to the ideal mixing these two states should have
comparable widths \cite{Cohen:2004gu}, while experimentally they are differ
substantially. The discussion of masses and decay widths of the $N^{\ast}$
states under assumption that they correspond to the Roper and $N^{\ast}(1710)$
done in Ref.~\cite{Pakvasa:2004pg} still indicates that it is impossible to
match the mass splittings with the observed branching ratios for these two
resonances even for arbitrary mixing. Whether any different assignment of the
diquark $N^{\ast}$ states would be compatible with the decay patterns
deserves a separate study.

In this short note we have argued that due to the smallness of the reduced
matrix elements of $\overline{10}\rightarrow8$, which is natural in chiral
soliton models, mixing with other SU(3) representations has to be taken into
account. Unfortunately, for the time being we can only speculate which mixing
scenario is phenomelogically impossible, allowed or desired. Here we have
examined a possibility that antidecuplet mixes with the Roper octet. Mixing of
Roper and the ground state octets is presumably very small. Indeed the first
order GMO mass formulae work very well for the ground state octet so there is
almost no space for additional mixing.

\noindent{\bf Acknowledgements}
This paper is based on a common work with Maxim Polyakov and
Klaus Goeke. I would like to thank the organizers of the
workshop "Excited QCD" for stimulative and creative atmosphere.

\end{document}